\documentclass[10pt]{IEEEtran}
\usepackage{amsmath,amsfonts, amsthm, amssymb}
\usepackage{algorithmic}
\usepackage{algorithm}
\usepackage{array}
\usepackage[caption=false,font=normalsize,labelfont=sf,textfont=sf]{subfig}
\usepackage{textcomp}
\usepackage{stfloats}
\usepackage{url}
\usepackage{verbatim}
\usepackage{graphicx}
\usepackage{cite}
\usepackage{caption}
\usepackage[letterpaper, top=0.7in, bottom=0.9in, textwidth=7.26in]{geometry}
\usepackage{xcolor}
\usepackage{enumitem}
\usepackage{thmtools}
\usepackage{centernot}
\usepackage{nopageno}

\declaretheoremstyle[
  headfont=\bfseries, 
  bodyfont=\normalfont,
]{boldstyle}

\declaretheorem[style=boldstyle, name=Theorem]{theorem}

\declaretheorem[style=boldstyle, name=Definition]{definition}

\begin{document}

\title{Lossy Semantic Communication for the Logical Deduction of the State of the World}

\author{Ahmet~Faruk~Saz,~\IEEEmembership{Student Member,~IEEE,}
        Siheng~Xiong,~\IEEEmembership{Student Member,~IEEE,}
        and~Faramarz~Fekri,~\IEEEmembership{Fellow,~IEEE}
}

\maketitle

\begin{abstract}

In this paper, we address the problem of lossy semantic communication to reduce uncertainty about the State of the World (SotW) for deductive tasks in point to point communication. A key challenge is transmitting the maximum semantic information with minimal overhead suitable for downstream applications. Our solution involves maximizing semantic content information within a constrained bit budget, where SotW is described using First-Order Logic, and content informativeness is measured by the usefulness of the transmitted information in reducing the uncertainty of the SotW perceived by the receiver. Calculating content information requires computing inductive logical probabilities of state descriptions; however, naive approaches are infeasible due to the massive size of the state space. To address this, our algorithm draws inspiration from state-of-the-art model counters and employs tree search-based model counting to reduce the computational burden. These algorithmic model counters, designed to count the number of models that satisfy a Boolean equation, efficiently estimate the number of world states that validate the observed evidence. Empirical validation using the FOLIO and custom deduction datasets demonstrate that our algorithm reduces uncertainty and improves task performance with fewer bits compared to baselines.
\end{abstract}

\begin{IEEEkeywords}
First-order logic, semantic communication, deduction, content-information, model counting

\end{IEEEkeywords}

\section{Introduction}

\thispagestyle{empty}

\IEEEPARstart{A}{s} the world progresses towards a future enriched by an array of Internet of Things (IoT) devices, the demand for ultra-reliable, low-latency, power- and bandwidth-efficient, context-aware connectivity becomes crucial. This need is prominent in emerging technologies such as spatial computing, autonomous vehicles, drones, smart cities, AI-driven healthcare, and Industry 4.0. Semantic communication, which prioritizes the relevance, significance, and context of information—not just the data itself—addresses this challenge effectively  \cite{Gund_Yen, Park2022, Hashash}.

Reasoning is crucial in semantic communication networks, supporting applications ranging from automated truth verification to the prevention of misinformation spread, data connectivity and retrieval, financial modeling and risk analysis, as well as privacy-enhancing technologies for data sharing and optimized agriculture. Logical reasoning techniques \cite{xiongtilp, yang2022temporal, xiong2024teilp, xiong2024large} such as deduction (applying principles to data points for inferring specific outcomes) can greatly benefit from a deep and nuanced understanding of the current environment, including the relationships and dependencies within it. However, in scenarios involving bandwidth limitations, privacy concerns, or multi-purpose surveillance, the central server may not wish to disclose the specific application to a sensor. In such cases, where the sensor is unaware of the central decision-maker’s specific task and communication resources are limited, the objective shifts to transmitting the most informative message possible about the environment. This is, on the average, expected to reduce the server's uncertainty regarding the present context or situation, leading to a more robust decision.

Therefore, in this paper, we explore a First-Order Logic (FOL)-based semantic communication framework that is not only suitable for logical deductive decision-making but also effectively transmits the most semantic information \cite{saz2024model} about the state of its environment. To that purpose, Carnap introduced a FOL-based approach \cite{c1} to semantic communication, representing the State of the World (SotW) in a finite universe using FOL. His state descriptions define the truth value of each predicate for all individuals involved, offering a method to quantify semantic information through inductive probability distributions. Building on Carnap's work, Hintikka \cite{c10} and his students extended this framework to accommodate universes with infinitely many individuals, addressing issues of inductive generalization and integrating logical quantifiers. A persistent challenge in these frameworks, which led to their abandonment by the engineering community, is the computational complexity of managing large state spaces. Our work builds on the ideas of Carnap and Hintikka, introducing a new paradigm that efficiently handles exponentially large state spaces using advanced model counting tools.

Thus, more formally, this paper addresses the challenge of developing a FOL-based lossy semantic communication paradigm that aims to transmit the most information about the environment, despite the transmitter's lack of task-specific knowledge and within a constrained bit budget. The goal is to reduce the receiver's uncertainty about the world state and enable the most accurate possible deduction of the SotW, leading to high deductive task performance.

The key contributions of this work are summarized as follows:

\textbf{Novel Semantic Communication Framework for Logical Reasoning Tasks:} We develop a custom ranking algorithm for point-to-point (P2P) communication, utilizing rapid model counters to select the most content-informative message for transmission, which prioritizes the preservation of semantic information for logical reasoning.

\textbf{Suitability for Logical Deduction Tasks:} FOL-based semantic communication framework presented in this manuscript is tailored towards logical reasoning tasks such as deduction. 

\textbf{Efficient Model Counting:} We address the challenge of the exponentially large state space under FOL representation by reframing the problem as a Boolean satisfiability model counting task, employing a state-of-the-art tree-based counter \cite{korhonen_et_al} for computationally efficient and rapid model counting, as well as the calculation of inductive logical probabilities.

\textbf{Convergence to Optimal SotW:} We demonstrate that as the accumulation of new observations approaches infinity, the reduction in semantic content uncertainty enables the receiver to learn the true State of the World (SotW). Consequently, any deductive task at the edge node converges to its optimal solution.

\textbf{Empirical Validation:} We conducted experiments using the FOLIO and custom deduction datasets. Our results show that nodes can improve their understanding of the SotW  with significantly reduced communication compared to baseline methods.

In the next section, we discuss related works in the area of semantic communication.

\section{Related Works}

Traditional logic-based methods, such as Carnap’s Semantic Information Theory and Floridi's Strongly Semantic Information Theory, have faced criticism for computational inefficiency, theoretical constraints, and their lack of a probabilistic structure. In response, early semantic communication efforts beyond 5G explored alternative solutions, utilizing techniques like simulated annealing, Deep Joint Source-Channel Coding (D-JSCC), distributed functional compression, autoencoders, deep-learning based semantic communication (DeepSC), and Graph Neural Networks (GNNs) \cite{yashas, GLi, Bennis2022b, gul-yen, Qin2021SemanticCP, aguerri2019distributed, zaidi2020distributed, xu2020acceleration, gupta2020training, krouka2021communication, Xie2020DeepLE, Shao2022ATO}. While these methods achieved considerable success on the tasks they're designed for, they are not suitable for logical reasoning which requires explicit representation in the form of First Order Logic \cite{yang2023harnessing}.

In contrast, our framework is specifically designed to alleviate the communication burden in logical deduction tasks and enhance scalability within computationally intensive environments. Furthermore, our approach offers a human-understandable format and broad applicability across multiple data types, essential for effective decision-making scenarios.

In the next section, we will shift our focus to background information and problem setup.

\section{Background on FOL Representation of the World State and
Measure of Semantic Uncertainty}

In this section, we start by introducing the primer necessary before moving onto the communication framework.

\subsection{Representation of the State of the World via FOL}

We define a finite world \(\mathcal{W}\) with possible states \(\mathcal{S}_{W_F} = \{s_1, s_2, \dots, s_v, \dots, s_{2^{|\mathbf{P}| \times |\mathbf{E}|^2}}\}\), where each state \(s_v\) is a First-Order Logic (FOL) sentence, using a language \(\mathcal{L}\). This language includes a set of dyadic predicates \(\mathbf{P}\), entities \(\mathbf{E}\), quantifiers \(\forall\) and \(\exists\), and a complete set of logical operators \(\mathbf{O} = \{\land, \lor, \neg\}\). In a finite world \(\mathcal{W}_F\), the state can be precisely described as follows (From this point on, we denote the index set of a set $X$ with $\mathcal{I}_x$):

\begin{definition}[\textbf{State Description}]
A \textit{state description} \( s_v \) in \(\mathcal{W}_F\) is:
\begin{equation}
    s_v = \bigwedge_{i \in \mathcal{I}_\mathbf{P}; \, j,k \in \mathcal{I}_\mathbf{E}} \sigma_{ijk} \cdot Pr_i(En_j, En_k),
\end{equation}
where \(\sigma_{ijk} \in \{+1, -1\}\) represents the presence (\(+1\)) or negation (\(-1\)) of the predicate \(Pr_i\) on entities \(En_j\) and \(En_k\), spanning all combinations to \(2^{|\mathbf{P}| \times |\mathbf{E}|^2}\). If \(Pr_i\) is monadic, it implies \(j = k\). These predicate-entity pairs might be like \(IsFriend(\text{Alice}, \text{Bob})\) indicating a relationship, or \(OwnsHouse(\text{Bob, Bob}) = OwnsHouse(\text{Bob})\) signifying ownership. 
\end{definition}

This setup allows the logical AND (\(\bigwedge\)) of all atomic sentences or their negations to represent any state in \(\mathcal{W}_F\). Moreover, any FOL sentence \(m\) in this world can be expressed as a disjunction of these state descriptions:

\begin{theorem}[\textbf{FOL Representation in Finite World \cite{c1}}] 
Any FOL sentence $m$ in \(\mathcal{W}_F\) can be represented as:
\begin{equation}
    m \equiv \bigvee_{i \in \mathcal{I}_{\mathcal{S}_{W_F}}} s_i,
\end{equation}
Examples are shown in Table \ref{ind_terms}.
\end{theorem}

\begin{table}[h]
\centering
\begin{tabular}{|>{\arraybackslash}p{8.25cm}|>{\arraybackslash}p{8.25cm}|}
\hline
\rule{0pt}{3ex}\textbf{Dyadic predicates: }$Eats$(Alice, Apple), $Owns$(Bob, Car), $Likes$(Alice, Bob), ... \\ \hline
\rule{0pt}{3ex}\textbf{State description: }$s_1$ $\equiv$ $\neg Eats$(Alice, Alice) $\land$ $Eats$(Alice, Apple) $\land \; (\neg$ $Eats$(Alice, Bob)) $\land \; (\neg$ $Eats$(Alice, Car))... (enumerations continue) ...\\ \hline
\end{tabular}
\vspace{1mm} 
\caption{FOL Representation in Finite World}
\vspace{-2mm} 
\label{ind_terms}
\end{table}

In an infinite world \(\mathcal{W}_I\), with a countably infinite entity set \(\mathbf{E}\) and \(e \rightarrow \infty\), specifying the state requires a more detailed formalization. We define a "Q-sentence" for such a scenario:

\begin{definition}[\textbf{Q-sentence}]
In \(\mathcal{W}_I\), a \textit{Q-sentence} \(Q_r(x,y)\) is a conjunction of predicate-entity pairs and their negations, structured as:
\begin{equation}
    Q_r(x,y) = \bigwedge_{i \in \mathcal{I}_\mathbf{P}} \left(\sigma_{i} \cdot Pr_i(x, y)\right) \land \left(\sigma'_{i} \cdot Pr_i(y, x)\right),
\end{equation}
where \(\sigma_{i}\) and \(\sigma'_{i}\) indicate the presence or negation of predicates between pairs \(x\) and \(y\) from \(\mathbf{E}\).

A more detailed representation involves "attributive constituents":
\end{definition}

\begin{definition}[\textbf{Attributive Constituent}]
An \textit{attributive constituent} \(Ct_n(x)\) in \(\mathcal{W}_I\) is:
\begin{equation}
    Ct_n(x) = \bigwedge_{u \in \mathcal{I}_\mathbf{U}; \, \exists y_j \in \mathbf{E}} \sigma_{u} \cdot Q_u(x, y_j),
\end{equation}
expressing all relationships \(x\) has within the world, with \(\sigma_{u}\) reflecting the presence or negation of each Q-sentence.
\end{definition}

We extend this to entire \(\mathcal{W}_I\) with "constituents":

\begin{definition}[\textbf{Constituent}]
Given set of attributive constituents $\mathcal{N}$, a \textit{constituent} \(C_w\) of width \(w\) in \(\mathcal{W}_I\) combines attributive constituents:
\begin{equation}
    C_w = \bigwedge_{n \in \mathcal{I}_\mathcal{N}; \, \exists x_i \in \mathbf{E}} \sigma_{n} \cdot Ct_n(x_i),
\end{equation}
encompassing the relational patterns of individuals within \(\mathcal{W}_I\).
\end{definition}

The FOL representation of any sentence \(m\) in \(\mathcal{W}_I\) can be decomposed into constituents:

\begin{theorem}[\textbf{Representation of FOL Sentences \cite{c10}}]
Given set of constituents $\mathcal{S}_{W_I}$, any FOL sentence \(m\) can be represented as:
\begin{equation}
    m \equiv \bigvee_{i \in \mathcal{I}_{\mathcal{S}_{W_I}}} C_i,
\end{equation}
Examples are shown in Table \ref{ind_terms_con}.
\end{theorem}

\begin{table}[h]
\centering
\begin{tabular}{|>{\arraybackslash}p{8.25cm}|>{\arraybackslash}p{8.25cm}|}
\hline
\textbf{Q-Predicates: }$Q_1$(Alice, Bob) $\equiv$ $Likes$(Alice, Bob) $\land$ $Likes$(Bob, Alice) $\land$ $\dots$\\ \hline
\textbf{Attributive-Constituents: }$Ct_1(\text{Alice}) \equiv$ $(\exists y)Q_{1}(\text{Alice, y}) \land$ $(\exists y)Q_{2}(\text{Alice, y}) \land$ $(\exists y)Q_{3}(\text{Alice, y})... \land$ $(\forall y)\{Q_{1}(\text{Alice, y}) \lor ... \}$ \\ \hline
\textbf{Constituents: }$C_m = (\exists x)Ct_1(x) \land \ldots \land (\exists x)Ct_m(x) \land (\forall x)(Ct_1(x) \lor \ldots \lor Ct_m(x))$ \\ \hline
\end{tabular}
\vspace{1mm} 
\caption{FOL Representation in Infinite World}
\vspace{-2mm} 
\label{ind_terms_con}
\end{table}

State descriptions and constituents function akin to basis vectors in a vector space, allowing the definition of a probability measure over them. The inductive logical probability of any sentence in a FOL language $\mathcal{L}$ is the sum of the probabilities of its constituents. Next, we discuss the probability measure definition for $\mathcal{L}$ using Carnapian and Hintikkan frameworks.

\subsection{Inductive Logical Probabilities of FOL Sentences}

In a FOL-based world \(\mathcal{W}\), we can define a probability measure \(\mathcal{P}\) over the foundational elements—either state descriptions $s_v$ 
or constituents $C_k$—termed generically as "states" (\(st\)) for simplicity. This probability measure \(\mathcal{P}\) functions as an inductive Bayesian posterior, built on empirical observations with a predefined prior distribution over these states.

\begin{definition}[\textbf{Inductive Logical Probability}]
For any state \(st\) and corresponding evidence \(e\) (observations) in finite or infinite \(\mathcal{W}\), the inductive logical probability or degree of confirmation \(c(st, e)\) that evidence \(e\) supports state \(st\) is:
\[
c(st, e) = \frac{p(st \land e)}{p(e)} = \sum_{t \in \mathcal{I}_c} \frac{n_c + \lambda(w_c)/w_c}{n + \lambda(w_c)},
\]
where \(\mathcal{I}_c\) indexes over set of states \( \mathcal{C}\) (where \( \mathcal{C}\) is $\mathcal{S}_{W_F}$ or $\mathcal{S}_{W_I}$), \(n\) is the total observation count, \(n_c\) is the count of observations confirming state \(st\), \(w_c\) reflects the state's weight, and \(\lambda(w_c)\) is the prior coefficient, which may be a function of the state \(st\)’s weight \(w_c\).
\end{definition}

\(\lambda(w_c)\), ranging from [0, $\infty$), influences the balance between empirical data and prior belief in the posterior. If a universal quantifier like \(\forall x \:\: HasDog(x)\) is present, it might be expanded to concrete instances such as \(HasDog(e_1) \land HasDog(e_2) \land \cdots\) to remove abstract variables and calculate probabilities. Conversely, existential quantifiers are handled by computing the probability for the negation, i.e., $c(\exists x \:\: HasCat(x), e) = 1 - c(\forall x \:\: \neg HasCat(x), e)$. 

Per Shannon’s classical information theory, the semantic information in observations (evidence \(e\)) should align with the reduction in uncertainty about world state \(st\), termed "semantic" surprise. However, inductive logical probabilities might not fully capture this element. Next, we explore an alternative measure to better gauge semantic surprise.

\subsection{Measure of Semantic Uncertainty about SotW}

In response to the need for a quantifiable metric of uncertainty reduction described in the previous section, Carnap introduced "cont-information" for assessing how observations affect an observer's understanding of the SotW. This metric measures the informational content of observations by evaluating how significantly they reduce uncertainty.

\begin{definition}
The \textit{cont-information} in a world \(\mathcal{W}\) quantifies the informativeness of an observation \(e\) to a specific observer by measuring the reduction in uncertainty about the SotW. It is defined as:
\begin{equation}
\text{cont}(m; e) = 1 - c(m, e) = 1 - p(m | e)
\end{equation}
where $m$ is any FOL sentence in $\mathcal{W}$ and \(e\) denotes the total observed evidence.
\end{definition}

This metric, in rough terms, simplifies to the proportion of state descriptions that an observation rules out. For example, in a finite world \(\mathcal{W}_F\) with two predicates in \(\mathbf{P}_F\) and two entities in \(\mathbf{E}_F\), there are 16 possible states. Assume uniform initial probabilities (\(\lambda(w_c) = w_c = 2^{|\mathbf{P}| \times |\mathbf{E}|}\)), giving each state an initial confirmation probability of \(\frac{1}{16}\). If an observation $(e_1 = Pr_1(en_1, en_2))$, is made by an observer (say Bob), this confirms certain states and invalidates others, refining \(\text{cont}(st_i; e_1)\) to \(\frac{7}{8}\) for states consistent with \(e_1\) and to \(1\) for inconsistent states. 

Now, consider another observer (say Alice), observing $(e_2 = Pr_1(en_1, en_2) \land Pr_1(en_2, en_1)$. This eliminates twelve states in total, giving more reduction in semantic surprise. Consequently, \( \text{cont}(st_i; e_2) \) for any of the remaining four states is \( \frac{3}{4} \), indicating a greater reduction in uncertainty compared to \( e_1 \) alone. If Bob in
addition makes the observation \(e_3 = Pr_1(en_2, en_1)\), \(e_1\) and \(e_3\) combined results in the same reduction in semantic surprise as Alice's, leading to \(\text{cont}(st_i; e_1 \land e_3) = \text{cont}(st_i; e_2) = \frac{3}{4}\).

This framework establishes a method to assess how different observations sequentially refine an observer’s certainty about the world's state, illustrating the dynamic and cumulative nature of evidence accumulation, and semantic surprise reduction. The next section will discuss a point-to-point communication setup to utilize this measure in practical scenarios.

\section{Point-to-Point Semantic Communication Framework}

\begin{figure}[!t]
\centering
\includegraphics[width=0.5\textwidth]{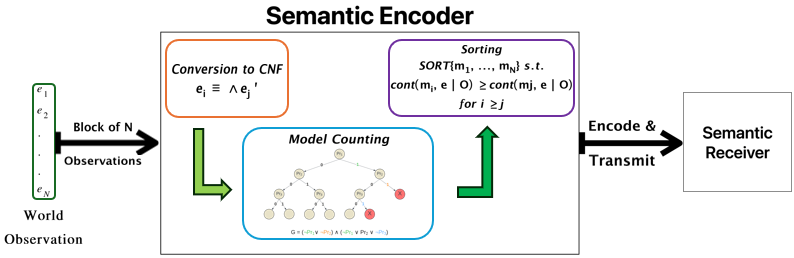}
\captionsetup{justification=centering}
\caption{Semantic communication framework}
\label{fig:my}
\end{figure}

We consider two nodes: a transmitter node \(N_{tx}\) and a receiver node \(N_{rx}\). The goal of \(N_{rx}\) is to deduce the state of the world as accurately as possible using information it receives from \(N_{tx}\) to increase it's success in a deductive task $\mathcal{T}$.  \(N_{tx}\) has a partial view of the SotW, as conveyed by its observations. The deductive task in question is to determine, given a mutually exclusive and jointly exhaustive hypothesis set \(\mathcal{H}\), the hypothesis \(h^*\) that best suits the population (i.e., the most consistent with the true state of the world).

The transmitter node \(N_{tx}\) makes observations \(e_i\), with its total evidence represented as \(\mathcal{E} = \bigwedge e_i\), a block of $N$ sentences. Its local perception of the State of the World (SotW) is a distribution over the set of states $\mathcal{C}$ denoted by \(\Pi_{tx}\). In absence of evidence, \(\Pi_{tx}\) is some prior distribution. Later, this distribution is updated to \(\Pi'_{tx}\) based on $\mathcal{E}$.

The receiver node \(N_{rx}\) also maintains a local perception of the SotW, a distribution \(\Pi_{rx}\), and aims to align his perception with the true distribution over world states \(\Pi^*\) as close as possible to improve his performance on a downstream logical reasoning task \(\mathcal{T}\). It must be emphasized, however, a perfect alignment may not be possible as \(N_{tx}\) may only have a partial view of the world through $\mathcal{E}$. \(\Pi_{rx}\) is the prior distribution, which is later updated to \(\Pi'_{rx}\) with information received.

\begin{figure}
    \centering
    \includegraphics[width=0.9\linewidth]{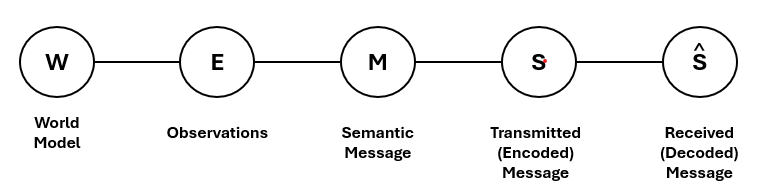}
    \caption{Graphical Model of Communication}
    \label{modelcom}
\end{figure}

Due to bandwidth limitations, \(N_{tx}\) can only transmit limited information under data transmission constraints. Initially, it is assumed both sender and receiver have a dictionary of all predicates and entities. The information to be communicated to receiver is strategically made to maximize the semantic content informativeness \(\text{cont}(m; \mathcal{E})\) of a message \(m\) (which is in the form of a FOL sentence), where $m$ is determined as a result of the following optimization problem:
\[
m^* = \text{argmax}_{m} \text{cont}(m; \mathcal{E} \mid \mathcal{O}),
\]

\noindent where $\mathcal{O}$ is the set of sentences that are already transmitted to \(N_{rx}\). In more formal terms, the optimization problem can be formulated as:

\begin{theorem}
\textbf{Lossy Content-Semantic Compression:} Consider lossy semantic communication of a message $M$ for a logical deduction task about the state of the world. The transmitted message $M$, on one hand, must maximize semantic content information regarding the state of the world, and on the other hand minimize the bit cost of transmitting the message. In other words, we must satisfy the following two conditions:

\begin{equation}\label{opteq}
\mathbf{M}^* = \text{argmax}_{c({\mathbf{m}}|\mathbf{e})} I_{\text{cont}}(\mathbf{M}; \mathcal{E} \mid \mathcal{O})
\end{equation}

\begin{equation}\label{opteq2}
\mathbf{S}^* = \text{argmin}_{p({\textbf{s}}|\textbf{m}^*)} I_{\text{shn}}(\mathbf{S}; \mathbf{M}^*)
\end{equation}
\end{theorem}

\noindent where \( I_{\text{shn}}(\textbf{S}; \textbf{M}) \) represents the Shannon mutual information, quantifying the amount of information \(\textbf{S}\) retains about \(\textbf{M}\) and $\mathcal{O}$ is the set of messages previously transmitted. \( I_{\text{cont}}(\textbf{M}; \mathcal{E}) = \sum c(m , e_i | \mathcal{O}) \times cont(m, e_i | \mathcal{O})\) defines the content-specific mutual information, indicating the extent to which the semantic significance of \(\mathcal{E}\) is preserved in \(\textbf{M}\). If there is a bit constraint on $M$, both constraints are jointly optimized and this may result suboptimal optimization of eqn. (\ref{opteq}), The communication process is represented in Fig. \ref{modelcom}. Upon receipt of message, \(N_{rx}\) updates its perception \(\Pi_{rx}\) to \(\Pi'_{rx}\) in order to more accurately reflect \(\Pi^*\). In the next section we discuss the relationship between semantic content uncertainty and success of deductive task \(T\).

\subsection{Analysis of the Relationship Between Semantic Content Uncertainty and Task Success}

We assume a hypothesis set \(\mathcal{H}'\), where each hypothesis \(h'\) is defined as a potential state \(st\) of the world \(\mathcal{W}\), where true state \(st^*\) corresponds to hypothesis \(h'^*\).

The discrepancy between any hypothesis \(h'\) and the true state \(h'^*\) is quantified by the loss function \(L(h', h'^*)\), with Bayes risk \(R_\Pi(h')\) representing the expected loss:
\[
R_\Pi(h') = E_{h' \sim \Pi, 'h^* \sim \Pi^*} [L(\mathbf{h}', \mathbf{h}'^*)].
\]
The loss function may be the symmetric set (Clifford) distance or a divergence between distributions such as Kullback-Leibler (KL) Divergence. Hintikka proved that \cite{c10} as the evidence accumulates, the semantic content informativeness \(\text{cont}(h'; \mathcal{E})\) trends toward 0 for all constituents $h'_i = C_i$ except for the true constituent $h'^* = C^*$ (i.e., the true SotW), i.e.,
\[
\lim_{e \to \infty} c(C_i; \mathcal{E}) = 1 \text{ if and only if } C_i = C^* \text{ and 0 otherwise.}
\]

Then, since \(N_{rx}\) updates its distribution \(\Pi'_{rx}\) based on message \(m\), which is in turn constructed based on evidence at \(N_{tx}\), if $|\mathcal{E}| \rightarrow \infty$, then \(\Pi'_{rx} \rightarrow \Pi^*\) and $cont(h';\mathcal{E}) \approx cont(h^*;\mathcal{E}) = 0$. Hence,  
\[
\lim_{e \to \infty} R_{\Pi'_{rx}}(h') \approx R_{\Pi^*}(h') = E_{h' \sim \Pi^*}[L(\mathbf{h}', \mathbf{h}'^*)] = 0.
\]

\subsection{Semantic Encoding Algorithm}

As previously discussed, the FOL-based semantic communication algorithm for deductive tasks at the edge suffer from exponentially large volume of the space. In fact, the number of possible states is given by $2^{|\mathbf{P}| \times |\mathbf{E}|^2}$, which is in the order of $\approx 10^{14} - 10^{15}$ for the well known Yet Another Great Ontology (YAGO) v4 dataset. Hence, calculating a brute-force inductive probability distribution over such dataset would be infeasible in terms of its time and space complexity.

In response to these challenges, we come up with an alternative solution based on state-of-the-art model counters for Boolean satisfiability (SAT) problems. In Boolean SAT problems, the goal is to determine if a set of logical constraints have any common solution, i.e., are satisfiable. If there exists at least one feasible assignment, the problem is SAT and otherwise non-SAT. Another interesting variation of this problem is the algorithmic set counting problems (ASC), which count the total number of models that satisfy given logical constraints.

\begin{figure}[h]
\centering
\includegraphics[width=0.3\textwidth]{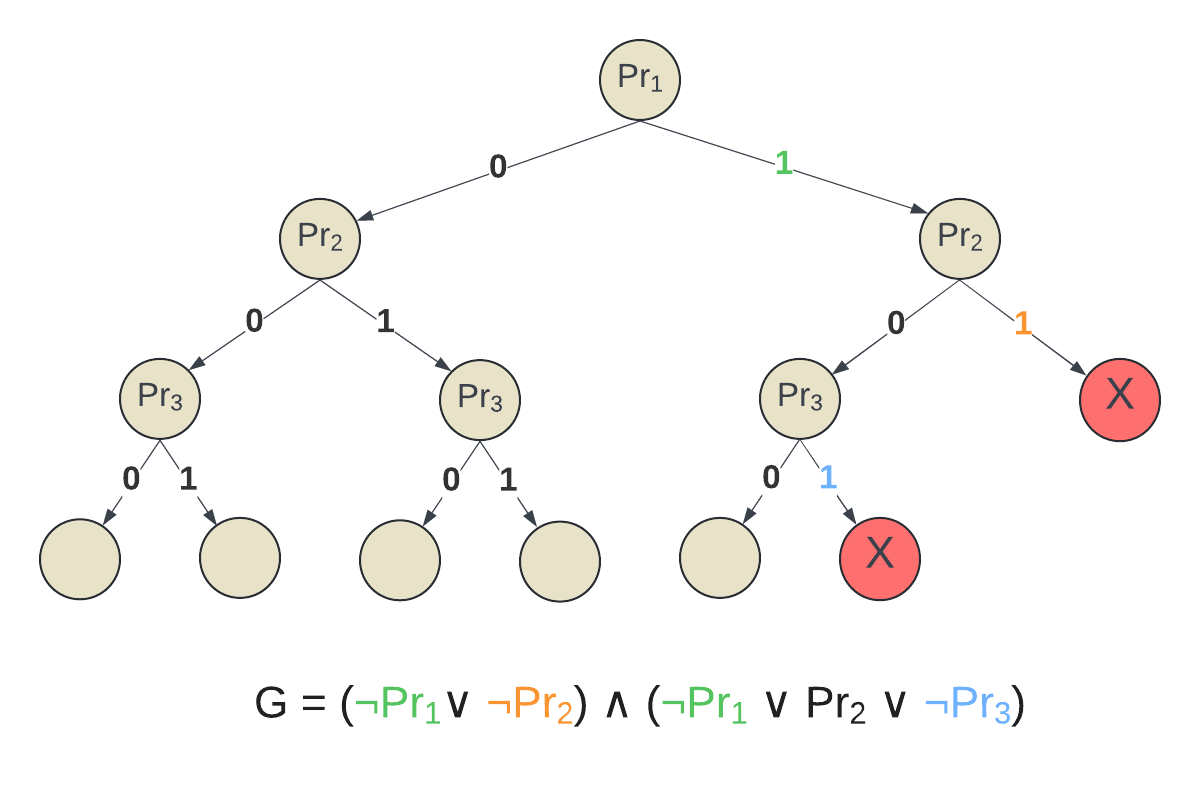}
\caption{SAT-Tree of proposition M for DPLL and SharpSAT-td}
\label{fig:yourlabel}
\end{figure}

Model counters use a binary tree structure, as shown in Fig. \ref{fig:yourlabel}, where each node is a decision point for a predicate with its branches representing true (1) or false (0) value assignments. This structure allows for recursive searching through the tree, employing logical constraints and heuristics to prune non-viable paths early in the search process, thereby reducing the computational overhead.

Given a FOL sentence (i.e., proposition) $M = (\neg Pr_1 \lor \neg Pr_2) \land (\neg Pr_1 \lor Pr_2 \lor \neg Pr_3)$, the assignment $Pr_1 = 1 \; \land \; Pr_2 = 1$ is eliminated early, without having to traverse its branches, hence reducing the computation time. The total number of SAT (satisfiable) assignments is calculated to be 5, the non-eliminated branches.

In the context of semantic content maximization, our communication algorithm selects a subset of FOL statements that yield the maximum reduction of content uncertainty at the receiver. As this reduction is contingent upon the number of satisfiable (SAT) assignments in an exponentially large space, a modified version of the SharpSAT-td tool, a state of the art Davis-Putnam-Logemann-Loveland solver for ASC problems.

During execution, FOL statements are initially processed into conjunctive normal form (CNF) via NLTK toolkit, and quantifiers are eliminated by translating them into equivalent grounded logical expressions. Then, SharpSAT-td counts the number of SAT models for each sentence and respective inductive probability is calculated. Then, the most informative statement is fed into a suitable encoder for compression (autoencoder, Huffman, Lempel-Ziv etc.), hence satisfying the bit-constraint. If another sentence is to be transmitted, the probabilities are recalculated to account for the first transmission, and the same process is repeated. A pseudo-code for the algorithm is provided below.

\begin{algorithm}
\caption{Semantic Communication Algorithm}
\begin{algorithmic}[1]
\REQUIRE A set $\mathcal{E}$ of $N$-FOL sentences
\ENSURE Subset of $M$-FOL sentences for transmission
\STATE $\mathcal{O} \gets \emptyset$ \COMMENT{Initialize an empty set to store transmitted $e_i$}
\STATE $\mathcal{D} \gets \mathbf{P'}, \mathbf{E'} \gets \textsc{Enumerate}(\mathbf{P}, \mathbf{E})$ \COMMENT{Create a dictionary of predicate-entity enumerations}
\STATE $K \gets M$
\WHILE{$K$ is not 0}
    \STATE $\mathcal{T} \gets \emptyset$ \COMMENT{Initialize a set to store model counts of $n_{e_i}$}
    \FOR{each sentence $e_i$ in $\mathcal{E}$}
        \STATE $\Phi(e_i) \gets \textsc{ComputeCNF}(e_i)$
        \STATE $n_{e_i} \gets$ \textsc{SharpSat-td}($\Phi(e_i)$)
        \STATE $\mathcal{T} \gets \mathcal{T} \cup (e_i, n_{e_i})$
    \ENDFOR
    \STATE $\mathcal{T'} \gets \textsc{Sort}(\mathcal{T}, \min_i c(e_i|\mathcal{O}, \textbf{e}))$
    \STATE $e^* \gets \{e_j \mid c({e_j} | \mathcal{O}, \textbf{e}) \leq c({e_k} | \mathcal{O}, \textbf{e}) \text{ for } \forall j,k \in \mathcal{T'}\}$
    \STATE Encode and Transmit $e^*$
    \STATE $\mathcal{O} \gets \mathcal{O} \cup e^*$, $K \gets K - 1$, $\mathcal{E} \gets \mathcal{E} \setminus \{e_i\}$
\ENDWHILE
\end{algorithmic}
\end{algorithm}

\section{Experiment Results}

We evaluated the performance of our semantic communication algorithm using the FOLIO dataset and a custom deduction dataset. The source code for the algorithm, its implementation, datasets, and experiments is available on GitHub \cite{sih_far}. FOLIO consists of 340 unique stories, each representing a distinct world \( W \), and we selected the 40 stories with at least 7 sentences for analysis. For each story, we identified the most informative subset of \( M \) sentences from a total of \( N \) sentences, where \( N \) varies between 7 and 10. We conducted experiments with subsets of size \( M = 1, 2, 3 \) and compared the semantic informativeness using the \textit{cont} metric. The reduction in uncertainty of the (transmitter) world perceived by the receiver as measured relative to the transmission of all \( N \) sentences, which represents 100\% reduction in uncertainty (the receiver has the same world model has the transmitter).

To validate the performance of our algorithm, we compare it against three baselines. The first is Huffman coding, the second is uniformly random selection, where a random subset of \( M \) sentences is chosen and transmitted, and the third is a GPT-based autoencoder for semantic compression. In the last method, we make use of the GPT-2 model, as downloaded from the Hugging Face to encode natural language stories into tokens and then, decode them back to their original form.

\begin{figure}
    \centering
    \includegraphics[width=0.8\linewidth]{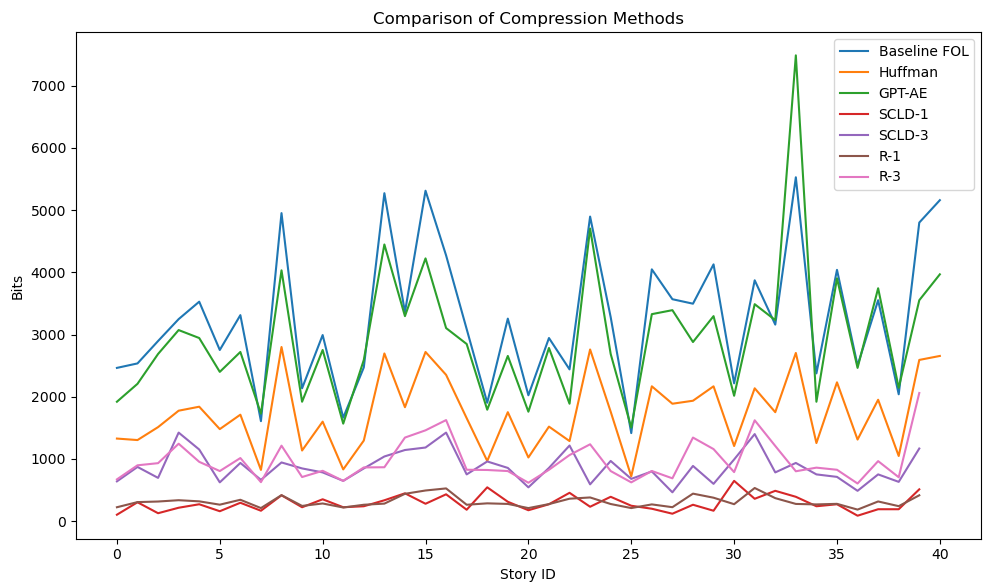}
    \caption{Comparison of bit costs of various methods}
    \label{fig:enter-label}
    \vspace{-0.25cm}
\end{figure}

\begin{table}[htbp]
\centering
\begin{tabular}{|c|c|c|c|c|}
\hline
\textbf{\# Sentences} & \textbf{SCLD Uncert.} & \textbf{R Uncert.} & \textbf{SCLD Cost} & \textbf{R Cost} \\ \hline
All & 100\% & - & 2420.1 bits & - \\ \hline
1 & 49.0\% & 37.4\% & 288.8 bits & 312.9 bits \\ \hline
2 & 72.7\% & 61.7\% & 600.6 bits & 646.1 bits \\ \hline
3 & 84.3\% & 76.4\% & 866.6 bits & 969.1 bits \\ \hline
\end{tabular}
\caption{Reduction in Uncertainty v.s. Comm. Cost}
\label{mytable}
\end{table}

\begin{figure}
    \centering
    \includegraphics[width=0.7\linewidth]{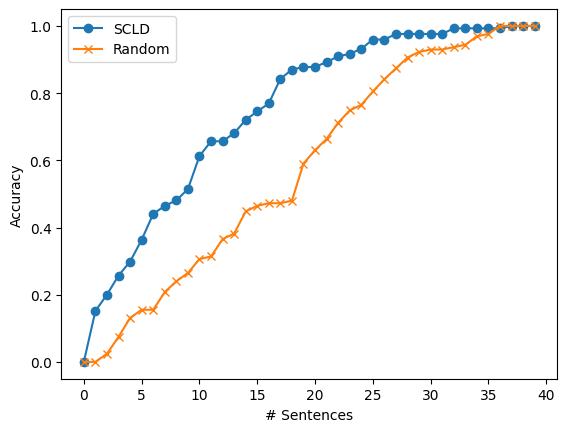}
    \caption{Success rate at hypothesis deduction task v.s. number of sentence transmissions}
    \label{fig:enter-label}
\end{figure}

Figure 3 shows the communication bit cost per story across different methods. The Baseline FOL, which transmits all FOL sentences, along with lossless Huffman coding and GPT-based autoencoding (GPT-AE), results in significantly higher communication costs while reducing the uncertainty at the same level. Although the GPT-based AE applies semantic compression to preserve meaningful features of the text for NLP tasks like summarization and allow perfect reconstruction of the text, it does not necessarily lead to a reduction in bit cost. In contrast, randomly transmitting 1 (R-1) or 3 (R-3) sentences results in lower bit costs but still falls short compared to our semantic communication for logical deduction (SCLD) method, which selects the most informative 1 (SCLD-1) or 3 (SCLD-3) sentences. On average, the random methods remain more costly and deliver less informative content.

These results are further elaborated in Table \ref{mytable}. The Baseline FOL, where all FOL sentences are transmitted (All), achieves the highest reduction in uncertainty (100\%) with a communication cost of 3280.78 bits. Our SCLD method outperforms the random (R) counterparts by 1.3X to 1.5X in uncertainty reduction, while using 1.08X - 1.11X less bits on average. For example SCLD-3 reduces uncertainty by 84.33\% compared to R-3's 76.40\%, while using 866.6 bits compared to R-3's 969.13 bits.

For the hypothesis deduction task, we created a custom deduction dataset of 10 stories, each consisting of 25 to 40 First-Order Logic sentences that describes the properties of the population of entities in terms of their attributes and relations. We assume the transmitting node observed this entire population (all sentences for each story). In this task, using the select sentences that are sent by the transmitter, the receiver must decide which hypothesis out of 8 disjoint hypotheses holds true for the population. We measure accuracy as the percentage of the population at the transmitter for which the receiver selected hypothesis can correctly predict the population properties, based on the received information. Figure 4 shows the results of these experiments across all stories as the number of transmitted sentences increases. The SCLD clearly chooses the sentences that improves significantly the quality of the deductive hypothesis selection compared to random transmissions.

In conclusion, the experiments confirm that the proposed semantic communication algorithm is highly effective in selecting and transmitting the most informative sentences, reducing uncertainty in a computationally efficient manner for deductive tasks at the edge under limited bandwidth, and outperforming its counterparts in both communication cost and uncertainty reduction.
\vspace{-0.5cm}

\bibliographystyle{unsrt} 
\bibliography{refs} 

\end{document}